\documentstyle[12pt, epsfig]{article}
\newlength{\dinwidth}
\newlength{\dinmargin}
\newcommand{\ba}{\begin{array}}
\newcommand{\ea}{\end{array}}
\newcommand{\be}{\begin{eqnarray}}
\newcommand{\ee}{\end{eqnarray}}

\newcommand{\ol}{\overline}
\newcommand{\wt}{\widetilde}
\newcommand{\wh}{\widehat}
\newcommand{\wl}{\widetilde\lambda}
\newcommand{\cl}{\cos(2\wl\pi)}
\newcommand{\ra}{\rightarrow}
\newcommand{\del}{\partial}
\newcommand{\half}{\frac{1}{2}}

\newcommand{\cF}{{\cal F}}
\newcommand{\cD}{{\cal D}}
\newcommand{\cL}{{\cal L}}
\newcommand{\cH}{{\cal H}}
\newcommand{\cT}{{\cal T}}

\def\vectt[#1,#2]{\left(%
\begin{array}{c} #1 \\ #2 \end{array} \right)}

\def\trivect#1{\trivectt[#1]}
\def\trivectt[#1,#2,#3]{\left(%
\begin{array}{c} #1 \\ #2 \\ #3 \end{array} \right)}

\newcommand{\bra}[1]{\left\langle\, #1\,\right|}
\newcommand{\ket}[1]{\left|\, #1\,\right\rangle}
\newcommand{\vac}{\ket{0}}

\def\nn{\nonumber \\}
\def\l{\left}
\def\r{\right}
\def\d{{\rm d}}

\newcommand{\gsim}{\mathrel{\mathop{\kern 0pt \rlap
  {\raise.2ex\hbox{$>$}}}
  \lower.9ex\hbox{\kern-.190em $\sim$}}}
\newcommand{\lsim}{\mathrel{\mathop{\kern 0pt \rlap
  {\raise.2ex\hbox{$<$}}}
  \lower.9ex\hbox{\kern-.190em $\sim$}}}

\begin{document}
\thispagestyle{empty} \addtocounter{page}{-1}
\begin{flushright}
SNUST 030101\\
{\tt hep-th/0301049}\\
\end{flushright} \vspace*{1.3cm}
\centerline{\Large \bf Rolling Tachyon with Electric and Magnetic
Fields} \vskip0.4cm \centerline{\Large \bf -- T-duality approach
--
 \footnote{Work supported in part by the BK-21 Initiative in
Physics (SNU Project-2), the KOSEF Interdisciplinary Research
Grant 98-07-02-07-01-5, and the KOSEF Leading Scientist Grant.}}
\vspace*{1.5cm}
\centerline{\bf Soo-Jong Rey${}^{a,b}$ {\rm and} Shigeki
Sugimoto${}^c$} \vspace*{1.0cm}
\centerline{
\it School of Physics \& Center for Theoretical
Physics} \vspace*{0.3cm}
\centerline{\it Seoul National University, Seoul 151-747 \rm
KOREA${}^a$}
\vspace*{0.5cm}
\centerline{ \it School of Natural Sciences, Institute for
Advanced Study} \vspace*{0.3cm} \centerline{\it Einstein Drive,
Princeton NJ 08540 {\rm USA}${}^b$}
\vspace*{0.5cm}
\centerline{ \it The Niels Bohr Institute} \vskip0.3cm
\centerline{\it Blegdamsvej 17, DK-2100, Copenhagen \O, \rm
DENMARK${}^c$} \vspace*{0.8cm}
\centerline{\tt sjrey@gravity.snu.ac.kr, sugimoto@nbi.dk}
\vskip2cm \centerline{\bf abstract} \vspace*{0.5cm}
We study the decay of unstable D$p$-branes when the world-volume
gauge field is turned on. We obtain the relevant D$p$-brane
boundary state with electric and magnetic fields by boosting and
rotating the rolling tachyon boundary state of a D$(p-1)$-brane
and then T-dualizing along one of the transverse directions. A
simple recipe to turn on the gauge fields in the boundary state is
given. We find that the effect of the electric field is to
parametrically enhance coupling of closed string oscillation modes
along the electric field direction and provide an intuitive
understanding of the result in the T-dualized picture. We also
analyze the system by using the effective field theory and compare
the result with the boundary state approach.


\baselineskip=18pt
\newpage

\section{Introduction}
Real-time dynamics of unstable D-brane has received considerable
attention recently \cite{sen1} - \cite{LNT}. Assuming that the
open string tachyon evolves homogeneously, Sen \cite{sen1} has
found that rolling of the tachyon down the potential hill toward
the closed string vacuum can be studied in terms of solvable
conformal field theory. Intriguingly, Sen's analysis indicates
that, at late time, the unstable D$p$-brane converts into a
pressureless tachyon matter \cite{sen2} localized on the
$p$-dimensional hyper-surface --- a result bearing potentially
interesting implications in the context of D-brane driven
cosmology \cite{cosmology}.

An interesting situation concerning rolling dynamics of the
tachyon is when gauge fields are excited on the world-volume of
decaying D-brane \cite{ishue,MW,MuSen,GHY}.
 On D$p$-brane world-volume, electric field
induces charge and current density of fundamental string, while
magnetic field induces charge and current density of
lower-dimensional D-brane. The physics we have in mind is, as a
D$p$-brane decays, how the fundamental strings and
lower-dimensional D-branes are distributed, and whether they can
be liberated out off the D$p$-brane and move freely in the ambient
space-time. A related question is how coupling of the decaying
D-brane to massless and massive closed string modes are modified
once the gauge field is turned on -- an issue of direct relevance
for non-commutative open string theory \cite{ncos}.

In this paper, we study the tachyon rolling in the electric and
magnetic field background on the world-volume of unstable D-brane.
The rolling tachyon boundary state with the electric field was
derived in \cite{MuSen} by carefully considering the effect of the
electric field to the boundary condition and the boundary
interaction. Here, instead, we utilize T-duality and Lorentz
transformation to turn on the constant gauge field and shed new
light on derivation and structure of the rolling tachyon boundary
state in the background gauge field. We find that the coupling to
massive closed string modes is affected by the electric field in
two notable ways. First, because of Lorentz contraction, coupling
to the Lorentz transformed modes is amplified hierarchically
--- unstable D-brane is more strongly coupled to higher closed
string modes. Second, the decay time-scale is Lorentz dilated,
prolonging the lifetime of the unstable D-brane. Combined
together, we conclude that the electric field imparts significant
modification to the decay of unstable D$p$-brane, especially, when
the electric field becomes critical. Intuitively, we interpret
this as a consequence of Lorentz enhancement of the D-brane energy
density in the T-dualized picture.

This paper is organized as follows. In section 2, we recapitulate
some of the results in \cite{sen1,sen2,MuSen,oksu} relevant for
foregoing analysis. In section 3, we show that tachyon rolling
dynamics in the gauge field background can be studied by an
elementary chain of maps involving Lorentz boost, rotation and
T-duality. We present a simple recipe for the corresponding
boundary state out of the boundary state for rolling tachyon. As a
corollary, we also present, beginning with a boundary state
describing spatial modulation of tachyon, an analogous recipe for
the boundary state describing tachyon modulation in world-volume
magnetic field background. In section 4, we study explicitly how
the coupling of the decaying D-brane to massless and massive
closed string states is modified by the presence of the
world-volume electric and magnetic field. In section 5, we analyze
the system using the effective field theory and compare the result
with those given in the previous sections.

\section{Boundary State of Rolling Tachyon}
In this section, we recapitulate aspects of rolling tachyon
boundary state, as analyzed in \cite{sen1,sen2,MuSen,oksu},
relevant for the analysis we will make in the subsequent sections.

Consider a D25-brane in bosonic string theory. In boundary
conformal field theory (CFT) description, classical dynamics of
the D25-brane world-volume is described by turning on an
appropriate boundary interaction. So, for the open string tachyon
$T(x)$ \footnote{We set $\alpha' = 1$ throughout this paper, so
that the tachyon mass-squared is given by $m^2 = - 1$.} rolling
down the world-volume potential hill toward the closed string
minimum,
\be T(x^0) \sim \lambda \cosh (x^0), \nonumber \ee
and the corresponding boundary interaction takes the form
\be \Delta S = \widetilde{\lambda} \oint \d \sigma \cosh
X^0(\sigma), \label{bdryaction} \ee
where $\widetilde{\lambda} = \widetilde{\lambda}(\lambda)$ is a
real parameter and $\sigma$ is the coordinate parameterizing the
world-sheet boundary. Making the Wick-rotation of $X^0 \rightarrow
i X$, dynamics of the Euclidean time $X(t, \sigma)$ described by
the world-sheet action
\be S_X =  \int \d t \oint \d \sigma \left[ (\partial_t X)^2 +
(\partial_\sigma X)^2 + \delta(t) \widetilde{\lambda} \cos
X(\sigma) \right] \label{worldsheetacgtion}\ee
turns out a solvable boundary conformal field theory
\cite{solvablecft1,solvablecft2,solvablecft3}. Making use of the
results of these works and Wick-rotating back to the Minkowski
time $X^0$, one can examine real-time rolling dynamics of the open
string tachyon.

The boundary state for a D25-brane with the boundary interaction
Eq.(\ref{bdryaction}) is expressed as
\be \vert {\rm D25} \rangle_T = \vert B \rangle_{X^0} \otimes \vert
N \rangle_{\bf X} \otimes \vert {\rm ghost} \rangle.
\label{D25}
\ee
Here, the latter two parts are the spatial and the world-sheet
ghost boundary-states for a flat D25-brane:
\be \vert N \rangle_{\bf X} &=&
\exp
\left( - \sum_{i=1}^{25}\sum_{n=1}^\infty {1 \over n}
\alpha^i_{-n} \overline{\alpha}^i_{-n} \right) \vert 0 \rangle,
\label{Neumann} \\ \vert {\rm ghost} \rangle &=& 
\exp
\left( - \sum_{n=1}^\infty (\overline{b}_{-n} c_{-n} + b_{-n}
\overline{c}_{-n} ) \right) (c_0 + \overline{c}_0) c_1
\overline{c}_1 \vert 0 \rangle, \nonumber \ee
while the first part $\vert B \rangle_{X^0}$ is the boundary state
describing real-time dynamics of the rolling tachyon.

In terms of the Wick-rotated variable $X$, the relevant boundary
state has been constructed
\cite{solvablecft1,solvablecft2,solvablecft3}. In case the
$X$-coordinate is compactified on a circle of self-dual radius
$R=1$, the boundary state is given by acting SU(2) rotation on the
unperturbed boundary state. The boundary state for non-compact
$X$-coordinate is then obtained by projecting onto zero-winding
subspace, and is given by
\be \vert B \rangle_{X} = \sum_{j=0,{1 \over 2}, 1, \cdots}
\sum_{m=-j}^{+j} D^j_{m, -m}(R) \vert j; m, m \rangle\!\rangle.
\ee
Here, $R$ is the SU(2) rotation matrix
\be R(\widetilde{\lambda}) = \left( \begin{array}{cc} \cos (\pi
\widetilde{\lambda}) & i \sin ( \pi \widetilde{\lambda}) \\
i \sin(\pi \widetilde{\lambda}) & \cos ( \pi \widetilde{\lambda})
\end{array} \right),
\ee
$D^j_{m,-m}(R)$ is the spin-$j$ representation matrix element for
the rotation, and $\vert j;m, m \rangle\!\rangle$ is the
Virasoro-Ishibashi state built over the primary state $\vert j; m,
m \rangle = \vert j, m \rangle \overline{\vert j, m \rangle}$.

The boundary state $\vert B \rangle_X$ is expressible in terms of
$X$-coordinate oscillators by facilitating the fact that the state
$\vert j, m \rangle$ belongs to the spin-$j$ representation of the
SU(2) current algebra defined by
\be J^\pm = \oint {\d z \over 2 \pi i} e^{\pm 2i X_R(z)}, \qquad
J^3 = \oint {\d z \over 2 \pi i} i \partial_z X_R (z). \nonumber
\ee
Explicitly,
\be \vert j, j \rangle = e^{ 2i j X (0) } \vert 0 \rangle \qquad
{\rm and} \qquad \vert j, m \rangle = N_{j,m} [J^-]^{(j-m)} \vert
j, j \rangle, \nonumber \ee
where $N_{j,m}$ is the normalization constant. Virasoro-Ishibashi
states, which preserves the diagonal part of the left- and the
right-moving Virasoro symmetries, are then constructed as
\be \left.\left.\vert 0; 0, 0 \right>\! \right> &=& \left( 1 + {1
\over 2} \alpha^2_{-1} \overline{\alpha}^2_{-1} + \cdots \right)
\vert 0 \rangle \nn \left.\left.\left\vert {1 \over 2}; \pm {1
\over 2}, \pm {1 \over 2} \right>\!\!\right> \right. &=& \left(1+
\alpha_{-1} \overline{\alpha}_{-1} + {1 \over 6} \left(
\alpha^2_{-1}\pm\sqrt\alpha_{-2} \right) \left(
\overline{\alpha}^2_{-1} \pm \sqrt{2}
\overline{\alpha}_{-2}\right) + \cdots \right) e^{\pm i X(0) } \,
\vert 0 \rangle \nn \vert 1; 0, 0 \rangle\!\rangle &=& \left(
\alpha_{-1} \overline{\alpha}_{-1} + {1 \over 2} \alpha^2_{-1}
\overline{\alpha}^2_{-1} + \cdots \right) \vert 0 \rangle \nn
\left\vert j; \pm j, \pm j \rangle\!\rangle \right.&=& \left( 1 +
\alpha_{-1} \overline{\alpha}_{-1} +{1 \over 2} \alpha^2_{-1}
\overline{\alpha}^2_{-1} + {1 \over 2} \alpha_{-2}
\overline{\alpha}_{-2} + \cdots \right) e^{\pm 2ij X(0)} \, \vert
0 \rangle   \qquad \qquad (j \ge 1) \nn \left.\left.\left\vert {3
\over 2}; \pm {1 \over 2}, \pm {1 \over 2} \right>\!\!\right>
\right. &=& \left({1 \over 6} \left( \alpha_{-2} \mp \sqrt{2}
\alpha_{-1}^2\right)\left(\overline{\alpha}_{-2} \mp \sqrt{2}
\overline{\alpha}^2_{-1} \right) + \cdots \right) e^{\pm i X(0)}
\, \vert 0 \rangle, \nonumber \ee
etc. Relative phase-factors among these Ishibashi states are
determinable by demanding the physics that the boundary state
represents an array of D-branes localized at $X = (2n+1) \pi$ when
$\widetilde{\lambda} = 1/2$ \cite{sen1}. Multiplying the relevant
matrix elements of the SU(2) rotation $R$:
\be D^j_{\pm j, \mp j}(R) &=& \left( i \sin (\pi
\widetilde{\lambda}) \right)^{2j} \qquad \quad (j=0, \, 1/2, \, 1,
\cdots) \nn D^1_{0\,,\,0}(R) &=& \cos (2 \pi \widetilde{\lambda})
\nn D^{3\over2}_{\pm{1 \over 2}, \mp{1 \over 2}} (R) &=& i
\sin(\pi \widetilde{\lambda}) \left( 3 \cos^2(\pi
\widetilde{\lambda}) - 1 \right) \nn D^2_{\pm 1, \mp 1}(R) &=& -
\sin^2(\pi\widetilde{\lambda}) \cos(2 \pi \widetilde{\lambda})
\nonumber \ee
to the Virasoro-Ishibashi states, and Wick-rotating back to the
$X^0$-coordinate, the sought-for boundary state is obtained as
\be \ket{B}_{X^0} &=& f(\wh x^0) \vac\nn
&+& g(\wh x^0)\,\alpha^0_{-1}\ol\alpha^0_{-1}\vac\nn
&+& h_1(\wh x^0)\,\alpha^0_{-2}\ol\alpha^0_{-2}\vac\nn
&+& h_2(\wh x^0)\,(\alpha^0_{-1})^2(\ol\alpha^0_{-1})^2\vac\nn
&+& h_3(\wh x^0)\,\l((\alpha^0_{-1})^2\ol\alpha^0_{-2}+
\alpha^0_{-2}(\ol\alpha^0_{-1})^2\r)\vac\nn &+& \cdots.
\label{expansion} \ee
The functions $f(x^0)$ is defined as
\be f(x^0) &=& \Big( 1 + e^{x^0} \sin (\widetilde{\lambda} \pi )
\Big)^{-1} + \Big( 1 + e^{- x^0} \sin (\widetilde{\lambda} \pi)
\Big)^{-1} - 1, \label{f} \ee
and the functions $g(x^0)$, $h_1(x^0)$, $h_2(x^0)$ and $h_3(x^0)$
are defined in terms of $f(x^0)$:
\be g(x^0) &=& 1 + \cos (2 \widetilde{\lambda} \pi) - f(x^0)
\label{g} \\
h_1(x^0) &=& {1 \over 2} \left(1 + \cos (2 \widetilde{\lambda}
\pi) \right) - \sin (\widetilde{\lambda} \pi) \cos^2
(\widetilde{\lambda} \pi) \cosh (x^0) - {1 \over 2} f(x^0) \nn
h_2(x^0) &=& 2 \sin (\widetilde{\lambda} \pi) \cos^2
(\widetilde{\lambda} \pi) \cosh (x^0) + {1 \over 2} f(x^0) \nn
h_3(x^0) &=& -i \sqrt{2} \sin (\widetilde{\lambda} \pi) \cos^2
(\widetilde{\lambda} \pi) \sinh (x^0). \label{h} \ee In the above
expressions, we have omitted the overall normalization factor
given by D$25$-brane tension $\cT_{25}$ for notational simplicity.
We will recover the normalization factor when we compute the
coupling to closed string states in section 4.

Notice that the function $f(x^0)$ exponentially vanishes at late
time, while the function $g(x^0)$ converges to a finite constant.
On the other hand, the functions $h_1(x^0)$, $h_2(x^0)$ and
$h_3(x^0)$ contain terms depending on either $\cosh(x^0)$ or
$\sinh(x^0)$, and hence blow up exponentially when $\vert x^0\vert
\rightarrow \infty$. As such, the tachyon matter couples
hierarchically strongly to higher-mass closed string states,
hinting that unstable D25-brane would preferentially populate
massive closed string modes rather than doing so massless
(graviton, dilaton, anti-symmetric tensor) modes \cite{oksu}.

\section{Tachyon Rolling in Constant Gauge Field Background}
In this section, starting from the result given in the previous
section, we construct the boundary state describing a D25-brane
with both rolling tachyon and homogeneous gauge fields turned on.
This is facilitated by making a chain of operations that leave
solvability of the conformal field theory intact.  To do so, we
will first consider a D25-brane, on whose world-volume all
excitations are set to zero except the rolling tachyon, and
compactify a spatial direction on a circle. We T-dualize it to a
D24-brane, and then boost the D24-brane rigidly along the
compactified spatial direction. Subsequently, we T-dualize back
along the boosted spatial direction. The final configuration we
obtain is a D25-brane, whose world-volume excitation involves both
rolling tachyon and {\sl constant} electric field. This argument
can be generalized to the cases with both electric and magnetic
fields by combining Lorentz boost and rotation in the T-dualized
picture. \footnote{ See for example \cite{DiVeLi} for a
comprehensive review of boosted and rotated D-brane boundary
states and their T-duality relations. }

\subsection{Turning on Constant Electric Field}
We begin with our prescription. Consider a D25-brane
\footnote{Extension to lower-dimensional D-branes is trivial. We
will phrase the prescription so that it is applicable to all
D$p$-branes $(p \ge 1)$.} in ${\bf R}^{1,25}$ and let the tachyon
field roll as in the previous section. We will compactify, say,
$x^1$-direction on a circle ${\bf S}_1$ of radius $R$, and wrap
the D25-brane on ${\bf S}_1$. In describing the open string
dynamics in terms of the boundary states, it is convenient to
decompose the world-sheet fields $X^a (z, \bar z)$ into the
left-moving part $X^a_{\rm L}(z)$ and the right-moving part
$\overline{X}^a_{\rm R}(\bar z)$, respectively.

We now perform a chain of maps, which retains the solvability of
the boundary conformal field theory. First, T-dualize the
$x^1$-direction, so that the world-sheet fields $X^{0,1}$ are
converted as
\be {\rm T-dual} \quad : \quad \left( \begin{array}{c} X^0_{\rm L} \\
X^1_{\rm L} \end{array} \right) &\rightarrow& \left(
\begin{array}{cc} +1 & 0 \\ 0 & +1 \end{array} \right) \left(
\begin{array}{c} X^0_{\rm L} \\ X^1_{\rm L} \end{array} \right)
\nn
\left( \begin{array}{c} \overline{X}^0_{\rm R} \\
\overline{X}^1_{\rm R} \end{array} \right) &\rightarrow& \left(
\begin{array}{cc} +1 & 0 \\ 0 & -1 \end{array} \right) \left(
\begin{array}{c} \overline{X}^0_{\rm R} \\ \overline{X}^1_{\rm R}
\end{array} \right). \label{tdual1} \ee
At the same time, the D25-brane is turned into an array of
localized D24-brane on the dual circle $\widetilde{\bf S}_1$ of
radius $\widetilde{R} = 2 \pi /R$. We then take the
decompactification limit $\widetilde{R} \rightarrow \infty$, and
isolate a localized D24-brane. In the boundary state description,
the process maps the $X^1$-part of the Neumann state
Eq.(\ref{Neumann}) into the following Dirichlet state:
\be \vert D \rangle_{X^1} = \exp \left( + \sum_{n=1}^\infty {1
\over n} \alpha^1_{-n} \overline{\alpha}^1_{-n} \right) \delta
(\widehat{x}^1) \, \vert 0 \rangle. \label{x1dirichlet}\ee
Notice that it involves Dirac delta-function of the zero-mode
operator, $\widehat{x}^1$. Next, we boost the D24-brane along
$x^1$-direction with velocity $e$. The world-sheet fields $X^0,
X^1$ are then mapped as
\be e-{\rm boost} \quad : \quad \left( \begin{array}{cc} X^0_{\rm L} \\
X^1_{\rm L}
\end{array} \right) &\rightarrow& \left( \begin{array}{cc} Y^0_{\rm L} \\
Y^1_{\rm L}
\end{array} \right) = \gamma \left( \begin{array}{cc}
1 & e  \\ e  & 1 \end{array} \right) \left( \begin{array}{c} X^0_{\rm L} \\
X^1_{\rm L}
\end{array}
\right) \nn \left( \begin{array}{cc} \overline{X}^0_{\rm R} \\
\overline{X}^1_{\rm R}
\end{array} \right) &\rightarrow& \left( \begin{array}{cc}
\overline{Y}^0_{\rm R} \\
\overline{Y}^1_{\rm R}
\end{array} \right) = \gamma \left( \begin{array}{cc}
1 & e \\ e & 1 \end{array}
\right) \left( \begin{array}{c} \overline{X}^0_{\rm R} \\
\overline{X}^1_{\rm R} \end{array} \right), \label{eboost} \ee
where $\gamma = 1/\sqrt{1 - e^2}$ denotes the Lorentz factor.
Notice that, the zero-mode constraint in Eq.(\ref{x1dirichlet})
for the localized D24-brane renders two elementary but significant
changes after the boost: because of Lorentz time dilation and
length contraction effects, we find that
\be \widehat{x}^0 \,\, &=&  \gamma^{-1}\, \widehat{y}^0 \,\,\, =
\sqrt{1 - e^2} \, \widehat{y}^0 \nn \delta(\widehat{x}^1) &=&
\delta \left(\gamma (\widehat{y}^1 + e \widehat{y}^0)\right) =
\gamma^{-1} \delta \left(\widehat{y}^1 + e \widehat{y}^0 \right).
\label{deltaftn}\ee
We then compactify the $y^1$-direction on a circle ${\bf S}'_1$ of
radius $R'$, and arrange an array of D24-branes. Finally,
T-dualize back along the $y^1$-direction, under which the
world-sheet fields transform as
\be {\rm T-dual} \quad : \quad \left( \begin{array}{c} Y^0_{\rm L} \\
Y^1_{\rm L} \end{array} \right) &\rightarrow& \left(
\begin{array}{cc} +1 & 0 \\ 0 & +1 \end{array} \right) \left(
\begin{array}{c} Y^0_{\rm L} \\ Y^1_{\rm L} \end{array} \right)
\nn
\left( \begin{array}{c} \overline{Y}^0_{\rm R} \\
\overline{Y}^1_{\rm R} \end{array} \right) &\rightarrow& \left(
\begin{array}{cc} +1 & 0 \\ 0 & -1 \end{array} \right) \left(
\begin{array}{c} \overline{Y}^0_{\rm R} \\ \overline{Y}^1_{\rm R}
\end{array} \right). \label{tdual2} \ee
Moreover, the zero-mode part Eq.(\ref{deltaftn}) in the boundary
state transforms as
\be \gamma^{-1} \delta (\widehat{y}^1 + e \widehat{y}^0) \vert 0
\rangle = \gamma^{-1} \sum_{n \in \bf Z} e^{i {n \over R'}
(\hat{y}^1 + e \hat{y}^0)} \vert 0 \rangle \,\,\,
\rightarrow \,\,\, \gamma^{-1} \sum_{m \in \bf Z} e^{i m
(\hat{y}^1_{\rm L} - \hat{y}^1_{\rm R} + e \hat{y}^0)
R'} \vert 0 \rangle, \label{dual}\ee
yielding D25-branes wrapped on a dual circle ${\bf S}'_1$ of
radius $\widetilde{R}' = 2 \pi /R'$.  Decompactify the
$y^1$-direction, $R' \rightarrow \infty$. Then, in
Eq.(\ref{dual}), only the $m=0$ term contributes to the boundary
state, and yields the Born-Infeld factor: $\sqrt{1 - e^2} \vert 0
\rangle$ \cite{BiCaDiVe}. The resulting configuration is a
D25-brane whose world-volume dynamics involves, in addition to the
rolling tachyon field $T(\gamma^{-1}y^0)$, a homogeneous electric
field $F_{01} = e$.

\subsection{Recipe for the Boundary State Construction}
\label{recipe}
Considerations of the previous subsection facilitates us an
elementary recipe for construction of the boundary state
describing tachyon rolling in a constant electric field
background. Begin with the D25-brane boundary state
Eq.(\ref{D25}) describing open string tachyon rolling dynamics.
Using the result Eq.(\ref{expansion}) and expanding in powers of
the string oscillators, the boundary state is expandable in
oscillator-level:
\be
\vert {\rm D25} \rangle_{T} &=&
\vert {\rm D25}\rangle^{\rm matter}_T
\otimes \vert {\rm ghost} \rangle \nn
\ee

\be \vert {\rm D25} \rangle^{\rm matter}_T &=& \vert B
\rangle_{X^0} \otimes \vert N \rangle_{X^1} \otimes_{i \ne 0, 1}
\vert N \rangle_{X^i} \nn &=& F(\wh x^0)\vert{0}\rangle +
G_{ab}(\wh x^0) \, \alpha^a_{-1} \overline{\alpha}^b_{-1} \vert 0
\rangle \nn  &+& H_{ab}(\wh x^0) \, \alpha^a_{-2}
\overline{\alpha}^b_{-2} \vert 0 \rangle + I_{abcd}(\wh x^0) \,
\alpha^a_{-1} \alpha^b_{-1} \overline{\alpha}_{-1}^c
\overline{\alpha}_{-1}^d \vert 0 \rangle + \cdots, \label{start}
\ee
where $\vert 0 \rangle$ denotes the Fock-space vacuum of
$\alpha^a, \overline{\alpha}^a$ oscillators.

\subsubsection{Turning on electric field}
As explained in the previous subsection, the
D25-brane boundary state describing rolling tachyon {\sl and}
electric field is then obtainable by applying sequentially the
chain of maps, Eqs.(\ref{tdual1}, \ref{eboost}, \ref{tdual2}), to
Eq.(\ref{start}). We find that the resulting boundary state is
given by the following replacement to Eq.(\ref{start}):
\be \vert 0 \rangle &\rightarrow& \gamma^{-1} \vert 0 \rangle \nn
x^0 &\rightarrow& \gamma^{-1} y^0 \nn \left( \begin{array}{c}
\alpha^0_{-n} \\ \alpha^1_{-n} \end{array} \right) &\rightarrow&
\Lambda^{-1}\left( \begin{array}{c} \beta^0_{-n} \\
\beta^1_{-n}
\end{array} \right) \nn
\left( \begin{array}{c} \overline{\alpha}^0_{-n} \\
\overline{\alpha}^1_{-n} \end{array} \right) &\rightarrow&
\Lambda \left( \begin{array}{c} \overline{\beta}^0_{-n} \\
\overline{\beta}^1_{-n} \end{array} \right),
\label{replace}
\ee
where
\be \Lambda = \gamma \left( \begin{array}{cc} 1 &  +e \\
 +e & 1 \end{array} \right) \qquad {\rm and} \qquad \Lambda^{-1} =
\gamma \left( \begin{array}{cc} 1 & - e \\ - e & 1 \end{array}
\right). \ee
Here, $\beta_{-n}^a$ and $\overline\beta_{-n}^a$ denote the
oscillators of $Y^a_L$ and $\ol Y^a_R$ after the T-duality
(\ref{tdual2}), respectively. The resulting boundary state is that
\be \vert{\rm D25} \rangle_{T, e}^{\rm matter} &=& F^e (\wh y^0)
\vert 0 \rangle + G^e_{ab}(\wh y^0) \, \beta^a_{-1}
\overline{\beta}^b_{-1} \vert 0 \rangle \nn &+& H^e_{ab}(\wh y^0)
\, \beta^a_{-2} \overline{\beta}^b_{-2} \vert 0 \rangle +
I^e_{abcd}(\wh y^0) \, \beta_{-1}^a \beta_{-1}^b
\overline{\beta}_{-1}^c \overline{\beta}_{-1}^d \vert 0 \rangle +
\cdots \nonumber\ee
where
\be F^e (y^0) &=& \gamma^{-1} \, F(\gamma^{-1} y^0) \nn G^e_{ab}
(y^0) &=& \gamma^{-1} \left({}^t \Lambda^{-1} G(\gamma^{-1} y^0)
\Lambda \right)_{ab} \nn H^e_{ab}(y^0) &=& \gamma^{-1} \left( {}^t
\Lambda^{-1} H(\gamma^{-1} y^0) \Lambda \right)_{ab} \nn
I^e_{abcd} (y^0) &=& \gamma^{-1} \left( {}^t \Lambda^{-1} \left(
{}^t\Lambda^{-1} I (\gamma^{-1} y^0) \Lambda \right)_{ac} \Lambda
\right)_{bd}, \ee
and so on. The physics behind the above transformation is clear.
Nonzero electric field induces the Lorentz time dilation effect,
and hence slowing down the tachyon-rolling time-scale.

\subsubsection{Turning on magnetic field}
It is straightforward to extend the above recipe to the boundary
state describing tachyon modulation in a constant magnetic field
background by replacing the $e$-boost Eq.(\ref{eboost}) by a
rotation. Begin with the D25-brane boundary state with tachyon
modulation along the $x^1$-direction:
\be
\vert {\rm D25} \rangle_T^{\rm matter}
&=& \vert B \rangle_{X^1} \otimes
\vert N \rangle_{X^2} \otimes_{i \ne 1,2} \vert N \rangle_{X^i}\nn
&=& F(\wh x^1)
 \vert 0\rangle
+ G_{ab}(\wh x^1) \, \alpha^a_{-1} \overline{\alpha}^b_{-1} \vert
0 \rangle \nn &+& H_{ab}(\wh x^1) \, \alpha_{-2}^a
\overline{\alpha}_{-2}^b \vert 0 \rangle + I_{abcd}(\wh x^1) \,
\alpha_{-1}^a \alpha_{-1}^b \overline{\alpha}_{-1}^c
\overline{\alpha}_{-1}^d \vert 0 \rangle + \cdots .
\label{tachyonmodulation} \ee
Then, after a chain of T-dual map $\rightarrow$ rotation in
$(1-2)$-plane $\rightarrow$ T-dual map, the boundary state
describing tachyon modulation in a constant magnetic field
background is given by the replacement in
Eq.(\ref{tachyonmodulation}) by
\be \vert 0 \rangle &\rightarrow& \widetilde{\gamma}^{-1} \vert 0
\rangle \nn x^1 &\rightarrow& \widetilde{\gamma}^{-1} y^1 \nn
\left(
\begin{array}{c} \alpha^1_{-n} \\ \alpha^2_{-n} \end{array}
\right) &\rightarrow& \Omega^{-1} \left( \begin{array}{c}
\beta^1_{-n}
\\ \beta^2_{-n} \end{array} \right) \nn \left(
\begin{array}{c} \overline{\alpha}^1_{-n} \\
\ol\alpha^2_{-n}
\end{array} \right) &\rightarrow& \Omega \left(
\begin{array}{c} \overline{\beta}^1_{-n} \\
\overline{\beta}^2_{-n} \end{array} \right), \ee
where $\widetilde{\gamma} = 1/ \sqrt{1 + b^2}$, and
\be \Omega = \widetilde{\gamma} \left( \begin{array}{cc} 1 & - b
\\ + b & 1 \end{array} \right) \qquad {\rm and} \qquad
\Omega^{-1}  = \widetilde{\gamma} \left( \begin{array}{cc} 1 & + b
\\ - b & 1 \end{array} \right). \ee
As a result, the sought-for boundary state is obtained as
\be \vert {\rm D25} \rangle_{T, b}^{\rm matter}
 &=& F^b (\wh y^1)
 \vert 0\rangle
 + G^b_{ab} (\wh y^1) \beta^a_{-1} \overline\beta^b_{-1} \vert
0 \rangle \nn &+& H^b_{ab} (\wh y^1) \beta^a_{-2}
\overline\beta^b_{-2} \vert 0 \rangle + I_{abcd}^b(\wh y^1) \,
\beta^a_{-1} \beta^b_{-1} \overline{\beta}_{-1}^c
\overline{\beta}_{-1}^d \vert 0 \rangle + \cdots, \ee
where
\be F^b(y^1) &=& \widetilde{\gamma}^{-1} F
(\widetilde{\gamma}^{-1} y^1) \nn G^b_{ab}(y^1) &=&
\widetilde{\gamma}^{-1} \left({}^t \Omega^{-1} G
(\widetilde{\gamma}^{-1} y^1) \Omega \right)_{ab} \nn H^b_{ab}
(y^1) &=& \widetilde{\gamma}^{-1} \left( {}^t \Omega^{-1} H
(\widetilde{\gamma}^{-1} y^1) \Omega \right)_{ab} \nn
I^b_{abcd}(y^1) &=& \widetilde{\gamma}^{-1} \left( {}^t
\Omega^{-1} \left( {}^t \Omega^{-1} I (\wt\gamma^{-1} y^1)
\Omega\right)_{bc} \Omega\right)_{ad} \ee
and so on.

\subsubsection{Turning on electric + magnetic fields}
\label{elemag} We can also extend the recipes given in the
previous subsections to the situation turning on both electric and
magnetic field. This is achieved by T-dualizing, boosting and
rotating in $(2+1)$-dimensional sub-space, and finally T-dualizing
back the system.

Suppose that the matter part of the boundary state we start with
is of the form
\begin{eqnarray}
\ket{\rm D25}^{\rm matter}_T &=& \Big( \vert B \rangle_{X^0}
\otimes \vert B \rangle_{X^1} \otimes \vert N \rangle_{X^2} \Big)
\otimes_{i\ne 0,1,2} \vert N \rangle_{X^i} \nn &=& F(\wh x^0,\wh
x^1)\ket{0}+G_{ab}(\wh x^0,\wh x^1)\,
\alpha^a_{-1}\ol\alpha^b_{-1}\ket{0} \nn &+& H_{ab}
(\widehat{x}^0, \widehat{x}^1) \, \alpha^a_{-2}
\overline{\alpha}^b_{-2} \vert 0 \rangle + I_{abcd}(\widehat{x}^0,
\widehat{x}^1) \, \alpha^a_{-1} \alpha^b_{-1}
\overline{\alpha}^c_{-1} \overline{\alpha}^d_{-1} \vert 0 \rangle
+ \cdots, \label{D25mat}
\end{eqnarray}
describing the tachyon field, modulated along the $x^1$-direction,
is rolling. Then the corresponding boundary state with both
electric field $e=F_{02}$ and magnetic field $b=F_{12}$ turned on
is given by the following replacement
\begin{eqnarray}
\ket{0}&\ra&\wt\gamma^{-1}\gamma^{-1}\ket{0}
\label{BIfac}
\\
x^0&\ra&\gamma^{-1}y^0
\label{eby1}\\
x^1&\ra&\wt\gamma^{-1}y^1-be\wt\gamma\, y^0
\label{eby2}\\
\trivect{\alpha^0_{-n},\alpha^1_{-n},\alpha^2_{-n}}&\ra&
\!\Lambda^{-1}\Omega^{-1}\!\trivect{\beta^0_{-n},\beta^1_{-n},\beta^2_{-n}}
\label{lorot1}
\\
\trivect{\ol\alpha^0_{-n},\ol\alpha^1_{-n},\ol\alpha^2_{-n}}&\ra&
\,\,\,\Lambda\Omega \,\,\,
\trivect{\ol\beta^0_{-n},\ol\beta^1_{-n},\ol\beta^2_{-n}},
\label{lorot2}
\end{eqnarray}
where
\begin{eqnarray}
\Lambda=
\left(
\begin{array}{ccccc}
\gamma&0&\gamma e'\\
0&1&0\\
\gamma e'&0&\gamma\\
\end{array}
\right) \qquad {\rm and} \qquad  \Omega= \left(
\begin{array}{ccccc}
1&0&0\\
0&\wt\gamma&-\wt\gamma b\\
0&+\wt\gamma b&\wt\gamma
\end{array}
\right),
\end{eqnarray}
in which
\begin{eqnarray}
\gamma=1/\sqrt{1-{e'}^2} \qquad {\rm and} \qquad
\wt\gamma=1/\sqrt{1+b^2}. \label{gamma}
\end{eqnarray}
Here, we set $e'=e/\sqrt{1+b^2}$. This redefinition of electric
flux is necessary, since the velocity of the D24-brane along
$y^2$-direction in the T-dualized picture is changed by the
rotation. Note that we have reproduced correctly the Born-Infeld
factor:
\begin{eqnarray}
\gamma^{-1}\wt\gamma^{-1}=
\sqrt{1-e^2+b^2}=\sqrt{-\det(\eta_{ab}+F_{ab})}. \label{BIfac2}
\end{eqnarray}
The recipe Eq.(\ref{eby1}) and Eq.(\ref{eby2}) follows from the
following relation in the T-dualized picture:
\begin{eqnarray}
\left(
\begin{array}{l}
y^0\\ y^1\\ y^2
\end{array}
\right)
=\Omega\Lambda
\left(
\begin{array}{l}
x^0\\ x^1\\ x^2
\end{array}
\right) \qquad {\rm with} \qquad x^2 = 0,
\end{eqnarray}
where the right-hand-side refers to a
D24-brane localized at $x^2=0$.

As a result, the matter part of the boundary state becomes
\begin{eqnarray}
\ket{\rm D25}_{T,e+b}^{\rm matter}&=& F^{e+b}(\wh y^0,\wh
y^1)\ket{0}+G^{e+b}_{ab}(\wh y^0,\wh y^1)\,
\beta^a_{-1}\ol\beta^b_{-1}\ket{0} \nn &+&
H_{ab}^{e+b}(\widehat{y}^0, \widehat{y}^1) \beta^a_{-2}
\ol{\beta}^b_{-2} \vert 0 \rangle + I^{e+b}_{abcd}(\widehat{y}^0,
\widehat{y}^1) \, \beta^a_{-1} \beta^b_{-1} \ol{\beta}^c_{-1}
\ol{\beta}^d_{-1} \vert 0 \rangle +\cdots, \nonumber
\end{eqnarray}
where
\begin{eqnarray}
F^{e+b}(y^0,y^1)&=&\gamma^{-1}\wt\gamma^{-1}
F(\gamma^{-1}y^0,\wt\gamma^{-1}y^1-be\wt\gamma\,y^0),
\label{Feb}\\
G^{e+b}_{ab}(y^0,y^1)&=&\gamma^{-1}\wt\gamma^{-1}
\left({}^t(\Omega\Lambda)^{-1}
G(\gamma^{-1}y^0,\wt\gamma^{-1}y^1-be\wt\gamma\,y^0)\,
\Lambda\Omega\,\right)_{ab}, \label{Geb}\\
H^{e+b}_{ab}(y^0,
y^1) &=& \gamma^{-1} \wt{\gamma}^{-1} \left({}^t(\Omega
\Lambda)^{-1} H (\gamma^{-1}y^0, \wt{\gamma}^{-1} y^1 - be\wt{\gamma}
y^0) \Lambda \Omega \right)_{ab}, \nn
I^{e+b}_{abcd}(y^0, y^1) &=&
\gamma^{-1} \wt{\gamma}^{-1} \left( {}^t (\Omega \Lambda)^{-1}
\left( {}^t (\Omega \Lambda)^{-1} I (\gamma^{-1} y^0,
\wt{\gamma}^{-1} y^1 - be \wt{\gamma} y^0) \Lambda \Omega
\right)_{bc} \Lambda \Omega \right)_{ad},\nonumber
\end{eqnarray}
and so on.

As a check, we apply our prescription to the boundary state of a
static D25-brane. The boundary state is given by the Neumann
state:
\be \ket{\rm D25}^{\rm matter}&=& \exp\left( - \sum_{n=1}^\infty
{1 \over n} \eta_{ab}\alpha^a_{-n} \overline{\alpha}^b_{-n}
\right) \vert 0 \rangle. \nonumber\ee
From Eqs.(\ref{BIfac}), (\ref{lorot1}) and (\ref{lorot2}), we
obtain the boundary state of a static D25-brane with electric and
magnetic field turned on
\be \ket{\rm D25}^{\rm matter}_{e,b}&=& \gamma^{-1}\wt\gamma^{-1}
\exp\left( - \sum_{k=1}^\infty {1 \over k}
\left({}^t(\Omega\Lambda)^{-1}\eta\,\Lambda\Omega\right)_{ab} \beta^a_{-k}
\overline{\beta}^b_{-k} \right) \vert 0 \rangle. \nonumber\ee
Using the relation Eq.(\ref{BIfac2}) and
\begin{eqnarray}
\left({}^t(\Omega\Lambda)^{-1}\eta\,\Lambda\Omega \right)_{ab}
&=&(1 - e^2 + b^2)^{-1} \left(
\begin{array}{ccccc}
-(1+e^2+b^2) &-2eb &-2e\\
-2eb&1-e^2-b^2 &-2b\\
+2e&+2b&1+e^2-b^2
\end{array}
\right)_{ab} \nn &=& \l(\frac{1-F}{1+F}\r)_{ab}, \label{FF}
\end{eqnarray}
where $F:=(F^a_{~\,b})=(\eta^{ac}F_{cb})$, we obtain the boundary
state as
\be \ket{\rm D25}^{\rm matter}_{e+b} = \sqrt{\det(1+F)}\,
\exp\left( -\sum_{n=1}^\infty {1 \over
n}\l(\frac{1-F}{1+F}\r)_{ab} \beta^a_{-n} \overline{\beta}^b_{-n}
\right) \vert 0 \rangle. \label{D25eb} \ee
The result is in perfect agreement with the Neumann boundary state
with constant gauge fields, as computed previously in \cite{CLNY}.

\subsection{Generalization to super-string theory}
It is straightforward to extend the derivation given in the
previous subsections for the problem of rolling tachyon on an
unstable D$p$-brane in super-string theories. If we use the
NSR-formulation, the Lorentz transformation and the T-duality map
are given in the same way as the bosonic string case. Therefore,
our recipe of turning on electric and magnetic fields given in the
previous subsections are equally applicable to the super-string
theories, with an understanding that the NSR fermionic oscillators
transform the same as Eqs.(\ref{lorot1}, \ref{lorot2}) under the
$e$-boost and the $b$-rotation. In particular, this renders that
the relations such as Eq.(\ref{Feb}) and Eq.(\ref{Geb}) continue
to hold. In the absence of the world-volume gauge field, explicit
form of the lower-level states in the rolling tachyon boundary
state in super-string theory was obtained in \cite{sen2,oksu}. For
the massless modes, compared with the bosonic string result, the
only change is in the functional form of $f(x^0)$. Adopting the
argument of \cite{sen2}, we can obtain the source of massless
closed string states in the rolling tachyon boundary state for
super-string by simply replacing the function $f(x^0)$ with
\begin{eqnarray}
f(x^0)=
\Big(1+e^{\sqrt{2}x^0}\sin^2(\widetilde{\lambda}\pi)\Big)^{-1}+
\Big(1+e^{-\sqrt{2}x^0}\sin^2(\widetilde{\lambda}\pi)\Big)^{-1}-1
\nonumber
\end{eqnarray}
in the corresponding results for bosonic string theory. It is then
evident that, applying the argument given in the previous
sub-sections, this is also the case even if we turn on constant
electric and magnetic fields.
\section{Coupling to Closed String States}
Based on the recipe prescribed in the previous section, we now
compute the boundary state representing the rolling tachyon with
electric and magnetic field and extract the coupling strength of
the unstable D-brane to lower-level closed string modes. In order
to extract the coupling, it is useful to recall that the closed
string wave function for the bosonic string is given as
\begin{eqnarray}
\ket{\Phi}=\l\{ T(\wh x)+h_{(ab)}(\wh x)
\alpha_{-1}^a\ol\alpha_{-1}^b +B_{[ab]}(\wh x)
\alpha_{-1}^a\ol\alpha_{-1}^b+ \phi_s(\wh x)
(\bar c_{-1}b_{-1}+c_{-1}\ol b_{-1}) \r\}c_1\ol c_1\vac+\cdots\nn
\end{eqnarray}
up to the oscillator-level (1,1) states, where ellipses denote
higher oscillator-level states. Here, $\phi_s(x)$ is Siegel's
dilaton field, and is related to the sigma model dilaton field
$\varphi(x)$ as
\begin{eqnarray}
\varphi(x)=\phi_s(x)-\frac{1}{2}h^a_{~a}(x), \label{dilaton}
\end{eqnarray}
which couples to the world-sheet curvature scalar and behaves as a
scalar under the linearized general coordinate transformation
\cite{CLNY}. The boundary state couples to the closed string field
through a term proportional to $\bra{\Phi}(c_0-\ol c_0)\ket{B}$
and hence it acts as a source for the closed string field in the
linearized equation of motion, $(Q_B+\ol Q_B)\ket{\Phi}=\ket{B}$,
where $Q_B$ is the BRST operator.

\subsection{Rolling Tachyon with Electric Field}
First, we demonstrate how to construct the boundary state for the
rolling tachyon with electric field. We start with the boundary
state representing the homogeneous tachyon rolling without gauge
fields given in Eq.(\ref{D25}) with Eq.(\ref{Neumann}) and
Eq.(\ref{expansion}). The boundary state with electric field
$e=F_{01}$ is obtained by the replacement summarized in
Eq.(\ref{replace}). The result is
\begin{eqnarray}
\ket{\rm D25}_{T,e} &=&\ket{B}_{Y^0}\otimes\ket{B}_{Y^1}\otimes_{i
\ne 0, 1}\ket{N}_{X^i} \otimes\ket{\rm ghost},
\label{tachelectric}
\end{eqnarray}
where
\be
\ket{B}_{Y^0}&=&
 f(\gamma^{-1}\wh y^0) \vert 0 \rangle
\nn
&+&
 g(\gamma^{-1}\wh y^0) \gamma^2
\left[ ( \beta^0_{-1} - e \beta^1_{-1} ) (\overline{\beta}^0_{-1}
+ e \overline{\beta}^1_{-1}) \right] \vert 0 \rangle \nn
&+& h_1(\gamma^{-1}\wh y^0) \gamma^2 \left[ (\beta^0_{-2} - e
\beta^1_{-2})(\overline{\beta}^0_{-2} + e \overline{\beta}^1_{-2})
\right] \vert 0 \rangle \nn
&+&
h_2(\gamma^{-1}\wh y^0) \gamma^4 \left[ (\beta^0_{-1} - e
\beta^1_{-1})^2 (\overline{\beta}^0_{-1} + e
\overline{\beta}^1_{-1} )^2 \right] \vert 0 \rangle \nn
&+& h_3 (\gamma^{-1}\wh y^0) \gamma^3
\left[ (\beta^0_{-1} - e \beta^1_{-1} )^2 (\overline{\beta}^0_{-2}
+ e \overline{\beta}^1_{-2} ) + (\beta^0_{-2} - e \beta^1_{-2}
)(\overline{\beta}^0_{-1} + e \overline{\beta}^1_{-1})^2 \right]
\vert 0 \rangle \nn
&+& \cdots, \label{timepart}
\ee
and
\be \vert B \rangle_{Y^1} = \gamma^{-1} \exp \left( -
\sum_{n=1}^\infty {\gamma^2 \over n} (\beta^1_{-n} - e
\beta^0_{-n} ) (\overline{\beta}^1_{-n}+e \overline{\beta}^0_{-n}
) \right) \vert 0 \rangle. \label{spacepart} \ee
Here, $f(x^0)$ and $g(x^0)$ are the functions given in
Eqs.(\ref{f}, \ref{g}), and $h_1(x^0), h_2(x^0), h_3(x^0)$ are the
functions given in Eq.(\ref{h}).
The resulting boundary state is consistent with that given in
\cite{MuSen}. Notice that, in Eqs.(\ref{timepart},
\ref{spacepart}), the power of the Lorentz factor $\gamma$ is
determined by the total number of $\beta,
\overline{\beta}$-oscillators.

\subsubsection{Tachyon Coupling}
The part of the boundary state that couples to the closed string
tachyon is given by $\vert 0 \rangle$, and is furnished by the
first term in the expansion of $\ket{\rm D25}_{T,e}$. Using
Eqs.(\ref{timepart},\ref{spacepart}), we find
\be \rho_{\rm tachyon}(y^0) = {\cal T}_{25} \gamma^{-1}
f(\gamma^{-1} y^0), \label{tachcoup} \ee
where we have reinstated dependence of the boundary state on the
D25-brane tension $\cT_{25}$. We see that the tachyon coupling
vanishes exponentially as $\vert y^0 \vert \rightarrow \infty$,
finding qualitatively the same behavior as the situation without
electric field. Time evolution of the tachyon coupling is plotted
in Fig.1 for various values of $\wt{\lambda}$. The result exhibits
that the tachyon coupling reaches vacuum value $0$ faster at
larger $\wt{\lambda}$, and is compatible with Sen's interpretation
that $\wt{\lambda}$ refers to the initial value of the open string
tachyon field. Notice that, when compared with the situation
without electric field, the decay process takes place slower by
the Lorentz factor $\gamma$, and weaker by the inverse Lorentz
factor $\gamma^{-1}$. The latter suppression factor is precisely
the Born-Infeld factor.

\begin{figure}[tb]
   \vspace{0cm}
   \epsfysize=6cm
   \centerline{\epsffile{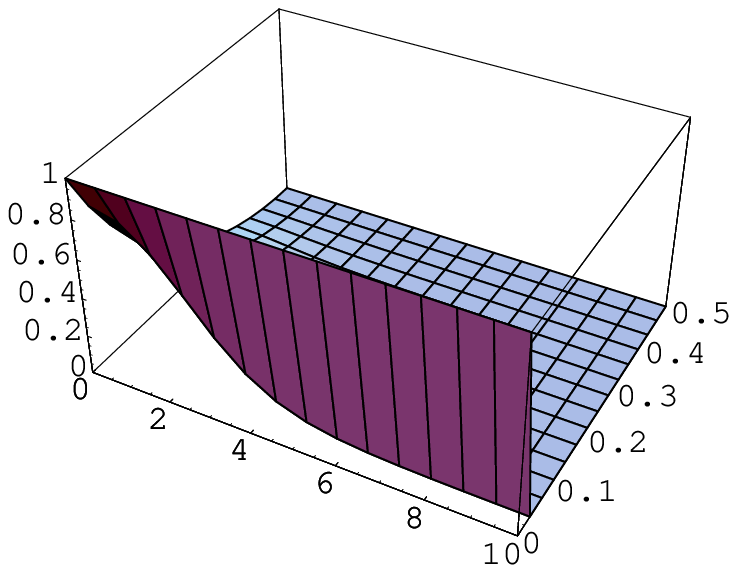}\hskip.5cm \epsffile{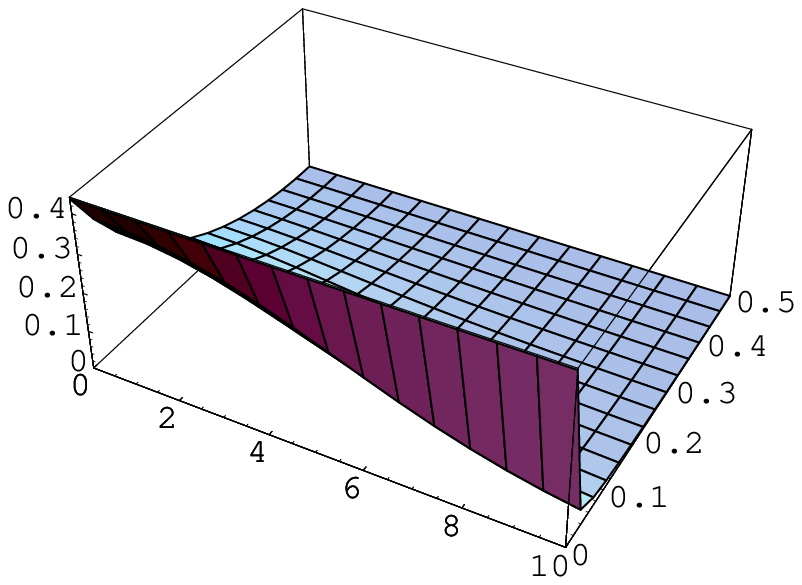}}
\caption{\sl Behavior of $\rho_{\rm tachyon} = \rho_{\rm dilaton}
=-T_{ii} (i \ne 0,1)$ (vertical axis in arbitrary unit) in
$y^0 = [0, 10]$ and $\widetilde{\lambda} = [0, 1/2]$
parameter space. The left is for $e=0$, while the right is for $e=0.9$.}
\label{fig01}
\end{figure}
\subsubsection{Graviton Coupling}
The boundary state that couples to the closed string graviton is
given by the energy-momentum tensor of the D25-brane. A
prescription of extracting energy-momentum tensor of the boundary
state was provided in \cite{sen2}. If the boundary state takes the
form
\be \ket{\rm D25}_{T,e}^{\rm matter} &=& F^e(\wh{y}^0) \vert 0
\rangle + G^e_{ab}(\wh{y}^0) \beta^a_{-1}\ol\beta^b_{-1} \vert 0
\rangle + \cdots, \label{mat} \ee
the energy-momentum tensor is given by
\be T^e_{ab}(y^0) := {{\cal T}_{25} \over 2} \left(G^e_{(ab)}(y^0)
- \eta_{ab} F^e(y^0) \right). \label{enemon} \ee
The second term in the right-hand side comes from the ghost
contribution, which couples to the trace of the graviton field,
$h^a_{~a}$, via the dilaton field relation Eq.(\ref{dilaton}). In
our case, we have
\be F^e(y^0) &=& \gamma^{-1} f(\gamma^{-1} y^0) \nn
 &=& \gamma (1 - e^2) f(\gamma^{-1} y^0) \nonumber \ee
and
\be G^e_{00}(y^0) &=& \gamma \, g(\gamma^{-1} y^0) + \gamma e^2 \,
f(\gamma^{-1} y^0) \nn &=& \gamma (1 + \cos (2 \widetilde{\lambda}
\pi) ) - \gamma (1 - e^2) f(\gamma^{-1} y^0),\nn G^e_{11}(y^0) &=&
- e^2 \gamma  \, g(\gamma^{-1} y^0) - \gamma f(\gamma^{-1} y^0)
\nn &=& - e^2 \gamma (1 + \cos (2 \widetilde{\lambda} \pi) ) -
\gamma (1 - e^2) f(\gamma^{-1} y^0),\nn G^e_{ii}(y^0) &=& -
\gamma^{-1} f(\gamma^{-1} y^0) \qquad \qquad \qquad \qquad (i=2,
\cdots, 25),\nn G^e_{01}(y^0)&=&-G^e_{10}(y^0)=e\gamma
(1+\cos(2\wl\pi)). \label{Gmat} \ee
Putting them together, we obtain non-vanishing components of the
energy-momentum tensor as
\be T^e_{00}(y^0) &=&+{\cT_{25}}
 \gamma \cos^2 (\widetilde{\lambda} \pi) \label{t00}\\
T^e_{11}(y^0) &=& -{\cT_{25}}
 e^2 \gamma \cos^2 (\widetilde{\lambda} \pi) - \cT_{25}
\gamma^{-1} f (\gamma^{-1} y^0) \label{t11}\\
 T^e_{ii}(y^0) &=&
- \cT_{25} \gamma^{-1} f(\gamma^{-1} y^0) \qquad \quad \qquad
\qquad (i = 2, \cdots, 25). \label{electricem} \ee
\begin{figure}[tb]
   \vspace{0cm}
   \epsfysize=6cm
   \centerline{\epsffile{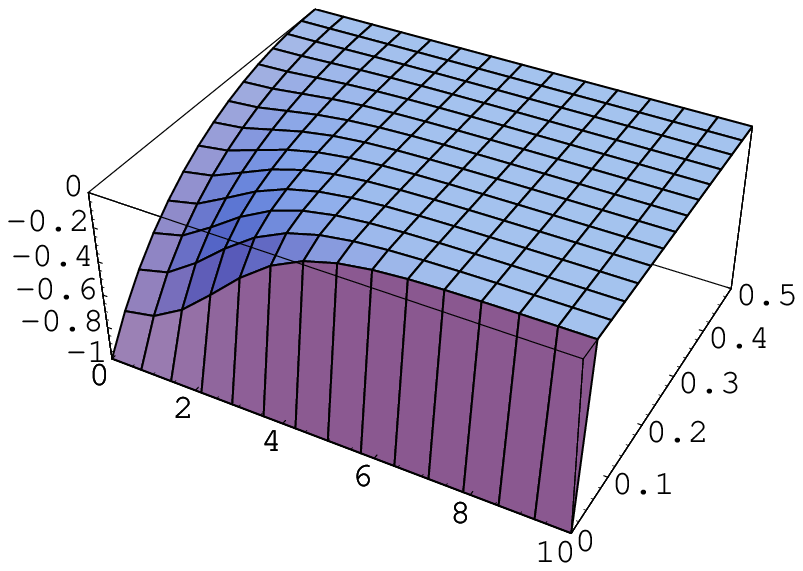} \hskip.5cm \epsffile{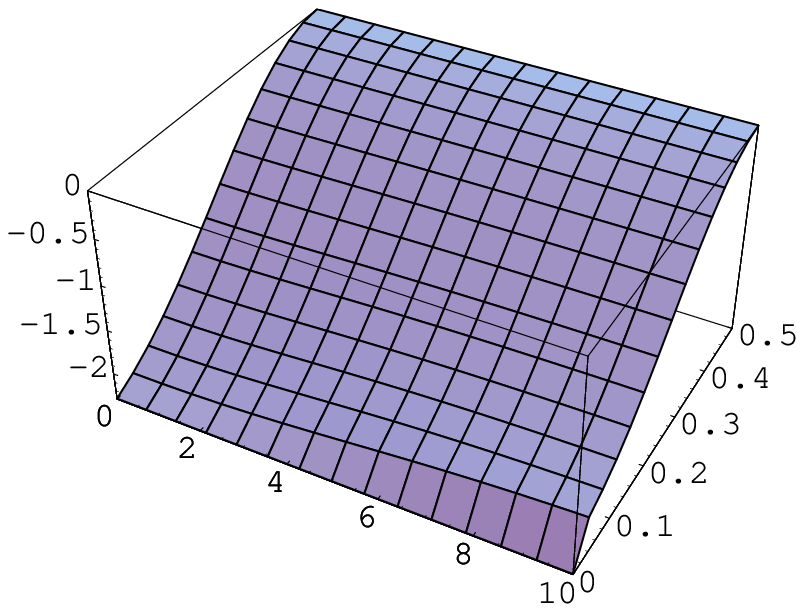}}
\caption{\sl Behavior of $T_{11}$ (vertical axis in arbitrary unit) in
$y^0 = [0, 10]$ and $\widetilde{\lambda} = [0, 1/2]$
parameter space.
The left is for $e=0$, while the right is
for $e=0.9$.} \label{fig02}
\end{figure}
The results are in agreement with those obtained in \cite{MuSen}
-- the energy-momentum tensor consists of a linear superposition
of the tachyon matter and the fundamental string fluid. We also
confirm that the energy density is independent of time, and hence
the energy-momentum tensor obeys the conservation condition
$\partial^a T_{ab} = 0$ trivially.
\subsubsection{Kalb-Ramond Coupling}
The boundary state that couples to the closed string Kalb-Ramond
field $B_{[ab]}$ is given by the anti-symmetric part of $G_{ab}^e$
in Eq.(\ref{Gmat}), and defines the fundamental string current
density tensor $Q^e_{[ab]}$:
\be Q^e_{01} := {{\cal T}_{25} \over 2} G^e_{[01]}={{\cal T}_{25}}
e\gamma \cos^2(\wl\pi). \label{Fdens} \ee
The result is again in agreement with that obtained in
\cite{MuSen}. We notice that the fundamental string current
density is a nonzero constant, and is seeded by the nonzero
electric field $e$. As such, it obeys the conservation condition
$\partial^a Q^e_{[ab]} = 0$ trivially.
As we will see in section 5,
the factor $\cos^2(\wt\lambda\pi)$ is intimately
related to the tachyon potential (as well as the world-volume
displacement field). Notice also that the coupling is enhanced by
the Lorentz factor $\gamma$.

{}From Eqs.(\ref{t00}, \ref{t11}, \ref{Fdens}), we observe that
these level-(1,1) couplings obey the following inequality
relations:
\be T^e_{00} \,\, \ge \,\, \left\vert Q_{01}^e \right\vert \,\,
\ge \,\, - T^e_{11} \qquad {\rm at} \qquad \vert y^0 \vert
\rightarrow \infty. \nonumber \ee
In the critical limit, $e \rightarrow 1$, we thus find that
\be T^e_{00} = - T^e_{11} = \left\vert Q^e_{01} \right\vert,
\nonumber \ee
yielding precisely a sort of BPS-type equation obeyed by the
fundamental closed string.

\subsubsection{Dilaton Coupling}
The boundary state that couples to the sigma model dilaton of the
closed string is given by
\be (c_0 + \overline{c}_0) (c_{-1} c_{1} + \overline{c}_{-1}
\overline{c}_1) \vert 0 \rangle_{\rm gh}. \nonumber \ee
Thus, the coupling of the decaying D25-brane to the closed string
dilaton is given by
\be \rho_{\rm dilaton}(y^0) = {\cal T}_{25} \gamma^{-1}
f(\gamma^{-1} y^0). \label{dilcoup} \ee
This is identical to the coupling to the closed string tachyon, so
exhibits the same asymptotic behavior. In particular, despite the
Lorentz time dilation and the Born-Infeld suppression effects, the
dilaton tadpole vanishes exponentially as $\vert y^0 \vert
\rightarrow \infty$.

\subsubsection{Massive State Coupling}
Coupling of the decaying D25-brane to a massive closed string
state can be read off from Eqs.(\ref{timepart}, \ref{spacepart}).
As is evident from Eq.(\ref{timepart}), power of the Lorentz
factor $\gamma$ is fixed by the number of $\beta,
\overline{\beta}$ oscillators for $0,1-$coordinates. Hence, at
$2n$-th level, among the boundary states created by these
$\beta$-oscillators, the state created by
\be \underbrace{\beta^0_{-1} \cdots \beta^0_{-1}}_{n_0}
\underbrace{\beta^{1}_{-1} \cdots \beta^1_{-1}}_{n_1}
\underbrace{\overline{\beta}^0_{-1} \cdots
\overline{\beta}^0_{-1}}_{n'_0} \underbrace{
\ol\beta^{1}_{-1} \cdots\ol\beta^1_{-1}}_{n'_1}
 \vert 0 \rangle \qquad {\rm with} \qquad n_0 +
n_1 =n'_0+n'_1= n \nonumber\ee
couples to the $2n$-th level closed string most strongly by the
factor $\gamma^{2n-1}$. We thus find that two important consequences
come about for the massive closed string state coupling of the
D25-brane once nonzero electric field is turned on. Recall that,
as was already noticed in \cite{MuSen,oksu}, the couplings to massive
closed string modes blows up as $\vert y^0 \vert \rightarrow
\infty$. This may be interpreted as an indication of strong
back-reaction effects. Once the electric field is turned on,
characteristic time-scale of the back-reaction is prolonged by the
(inverse) Born-Infeld factor $\gamma$, reflecting the familiar
Lorentz dilation effect. In addition, the overall coupling
strength is parametrically enhanced by $\gamma^{2n-1}$ at $2n$-th
level. This is to be contrasted with parametric suppression of the
tachyon and the sigma model dilaton couplings Eqs.(\ref{tachcoup},
\ref{dilcoup}) by the factor $\gamma^{-1}$.
\begin{figure}[tb]
   \vspace{0cm}
   \epsfysize=6cm
   \centerline{\epsffile{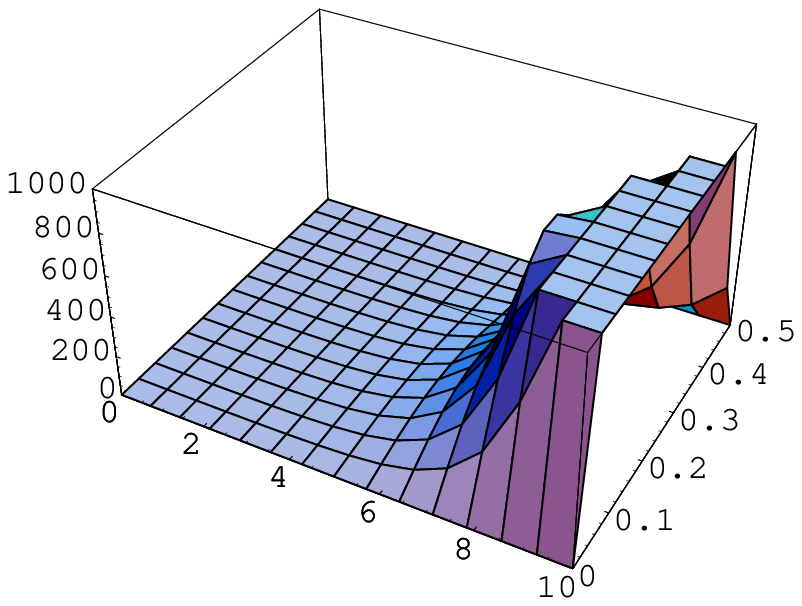} \hskip.5cm \epsffile{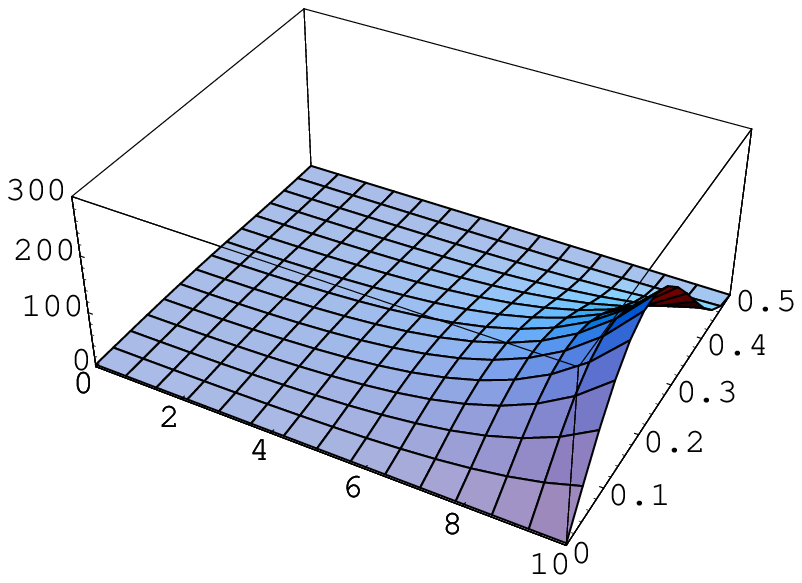}}
\caption{\sl Behavior of $h_2$ (vertical axis in arbitrary unit) in
$y^0 = [0, 10]$ and $\widetilde{\lambda} = [0, 1/2]$
parameter space. The left is for $e=0$, while the right is for
$e=0.9$.} \label{fig03}
\end{figure}
%
\subsection{Rolling Tachyon with Electric + Magnetic Field}
\label{elemagroll} Next, we consider turning on both electric and
magnetic fields on the D25-brane world-volume. We again start with
the boundary state Eq.(\ref{D25}) and then perform $e$-boost and
$b$-rotation. The matter part of Eq.(\ref{D25}) can be written as
Eq.(\ref{D25mat}), where the first two coefficient functions are
\begin{eqnarray}
F(x^0)=f(x^0), \qquad G_{00}(x^0)=g(x^0), \qquad
G_{ii}(x^0)=-f(x^0), \qquad (i=1,\cdots,25).
\label{elemagstart}
\end{eqnarray}
Here, $f(x^0)$ and $g(x^0)$ are the functions defined in
Eqs.(\ref{f}, \ref{g}).

We next turn on constant electric field $e=F_{02}$ and magnetic
field $b=F_{12}$. Following the prescription given in section
\ref{elemag}, we obtain the matter part of the corresponding
boundary state as
\begin{eqnarray}
\ket{\rm D25}^{\rm matter}_{T,e+b}&=& F^{e+b}(\wh y^0)
\ket{0}+G^{e+b}_{ab}(\wh y^0)\, \beta^a_{-1}\ol\beta^b_{-1}\vac+\cdots,
\end{eqnarray}
where the functions $F^{e+b}(y^0)$ and $G^{e+b}(y^0)$ are obtained
by inserting Eq.(\ref{elemagstart}) into the formulae
Eq.(\ref{Feb}) and Eq.(\ref{Geb}). Explicitly,
\begin{eqnarray}
F^{e+b}(y^0)&=&{1 \over \gamma\wt\gamma} f(\gamma^{-1}y^0),
\end{eqnarray}
and
\begin{eqnarray}
G^{e+b}_{00}(y^0)&=&+\gamma\wt\gamma(1+b^2)(1+\cl)) -{1 \over
\gamma \wt\gamma} f(\gamma^{-1}y^0),\nn
G^{e+b}_{11}(y^0)&=&+\gamma\wt\gamma \l(\frac{e^2b^2}{1+b^2}\r) (1
+ \cl) -{1 \over \gamma \wt\gamma}
\l(\frac{1-b^2}{1+b^2}\r)f(\gamma^{-1}y^0),\nn
G^{e+b}_{22}(y^0)&=&-\gamma\wt\gamma \l( \frac{e^2}{1+b^2} \r) (1
+ \cl) -{1 \over \gamma \wt\gamma}\left( \frac{1-b^2}{1+b^2}
\right) f(\gamma^{-1}y^0),\nn G^{e+b}_{ii}(y^0)&=&-{1 \over
\gamma\wt\gamma}f(\gamma^{-1}y^0) \qquad \quad
(i=3,\cdots,25),\nn
G^{e+b}_{01}(y^0)&=&+G^{e+b}_{10}(y^0)=\gamma\wt\gamma
eb\,(1+\cl),\nn
G^{e+b}_{02}(y^0)&=&-G^{e+b}_{20}(y^0)=\gamma\wt\gamma
e\,(1+\cl),\nn G^{e+b}_{12}(y^0)&=&-G^{e+b}_{21}(y^0)=
\gamma\wt\gamma \l( \frac{e^2b}{1+b^2} \r) (1 + \cl) +{1 \over
\gamma \wt\gamma} \l( \frac{2b}{1+b^2} \r) f(\gamma^{-1}y^0).
\nonumber
\end{eqnarray}

Thus the coupling to the closed string tachyon and the dilaton field
can be immediately read off as
\begin{eqnarray}
\rho_{\rm tachyon}(y^0)=\rho_{\rm dilaton}(y^0)=
\cT_{25}\frac{1}{\gamma\wt\gamma} f(\gamma^{-1}y^0).
\label{tacdila}
\end{eqnarray}

The energy-momentum tensor defined by Eq.(\ref{enemon}) is
\begin{eqnarray}
T^{e+b}_{00}(y^0)&=&+\cT_{25} \gamma\wt\gamma
(1+b^2) \cos^2 (\widetilde{\lambda} \pi),\nn
T^{e+b}_{11}(y^0)&=&+\cT_{25}\l(
\gamma\wt\gamma \l( \frac{e^2b^2}{1+b^2} \r)
\cos^2 (\widetilde{\lambda} \pi)
-{1 \over \gamma \wt\gamma} \l( \frac{1}{1+b^2} \r)
f(\gamma^{-1}y^0)\r),\nn
T^{e+b}_{22}(y^0)&=&-\cT_{25}\l(
\gamma\wt\gamma \l( \frac{e^2}{1+b^2}\r)
\cos^2(\widetilde{\lambda} \pi)
+{1 \over \gamma \wt\gamma}\l( \frac{1}{1+b^2}\r)
f(\gamma^{-1}y^0) \r),\nn T^{e+b}_{ii}(y^0)&=&-\cT_{25}{1 \over
\gamma\wt\gamma}f(\gamma^{-1}y^0), \qquad \qquad
(i=3,4,\dots,25),\nn
T^{e+b}_{01}(y^0)&=&T^{e+b}_{10}(y^0)={\cT_{25}} \gamma\wt\gamma
eb\, \cos^2 (\widetilde{\lambda} \pi). \label{Teb}
\end{eqnarray}

The fundamental string current density $Q_{[ab]}^{e+b} :=
\frac{1}{2} {\cal T}_{25} G_{[ab]}^{e+b}$ is obtained as:
\begin{eqnarray}
Q^{e+b}_{02}(y^0)&=&-Q^{e+b}_{20}(y^0)={\cT_{25}} \gamma\wt\gamma
e\, \cos^2 (\widetilde{\lambda} \pi),\label{Seb} \\
Q^{e+b}_{12}(y^0)&=&-Q^{e+b}_{21}(y^0)=\cT_{25} \l(\gamma\wt\gamma
\l( \frac{e^2b}{1+b^2}\r) \cos^2(\widetilde{\lambda} \pi) +{1 \over \gamma
\wt\gamma} \l( \frac{b}{1+b^2} \r) f(\gamma^{-1}y^0)\r). \nonumber
\end{eqnarray}

The result may be understood heuristically as follows. The
mutually orthogonal electric and magnetic fields give rise to
nonzero field momentum on the D$p$-brane world-volume (for
instance, for $p=3$, the field momentum is given by the Poynting
vector, whose magnitude is $eb$ and direction is perpendicular to
the plane spanned  by the electric and magnetic fields). As such,
non-vanishing components of the energy-momentum tensor
$T_{01}^{e+b}$ and the string current density tensor
$Q_{12}^{e+b}$ are interpretable as arising from rigid flow of the
fundamental string fluid. Indeed, it is possible to boost the
system along $1$-direction by
\be V = \frac{T_{01}^{e+b}}{T_{00}^{e+b}} = {eb \over 1 + b^2}
\qquad {\rm and} \qquad \Gamma = 1/ \sqrt{ 1 - V^2} \nonumber \ee
into an inertial `rest' frame, where the fluid is at rest. This
rigid flow is also expected from the relation Eq.(\ref{eby2}).
From the requirement that $e^2 \le (1 + b^2)$, we deduce that the
boost velocity $V$ never exceeds the speed of light. Note that the
magnetic field induces D$23$-brane density. However, as this
D$23$-brane is also unstable and will eventually evaporate, we
expect that the system behaves in a manner similar to that in pure
electric field at late time. Indeed, in the `rest' frame,
non-vanishing components of the energy-momentum tensor are
\be
\widehat{T}_{00}^{e+b} &=& +{\cal T}_{25} \cos^2 (\widetilde{\lambda} \pi)
\gamma \widetilde{\gamma} \left( {1 + b^2 \over \Gamma^2} \right)
\nonumber \\
\widehat{T}_{22}^{e+b} &=& - {\cal T}_{25} \cos^2
(\widetilde{\lambda} \pi) \gamma \widetilde{\gamma}
\left( {e^2 \over 1 + b^2} \right)
=-v_s^2\,\wh T_{00}^{e+b}, \nonumber
\ee
while that of the fundamental string current density tensor is
\be \widehat{Q}_{02}^{e+b} = {\cal T}_{25} \cos^2
(\widetilde{\lambda} \pi) \gamma \widetilde{\gamma} \left({e \over
\Gamma} \right) =v_s\,\wh T_{00}^{e+b}. \nonumber \ee
Here, we have denoted the sound velocity of the fundamental string
fluid as \be v_s^2 := \left\vert {\widehat{T}_{22} \over
\widehat{T}_{00}} \right\vert = {e^2 \over (1 + b^2)^2 - e^2 b^2},
\nonumber \ee
and suppressed $f(\gamma^{-1}y^0)$-dependent contributions, as
they drop out at late time or in the critical limit.

Since the gauge field strengths are constrained $e^2 \le (1 +
b^2)$, we find that the sound velocity cannot exceed the speed of
light
\begin{eqnarray}
v_s^2 \le {e^2 \over 1 + b^2} \le 1. \label{inequality2}
\end{eqnarray}
Therefore, it is now evident that the BPS-like
inequality comes about as
\be
\widehat{T}_{00}^{e+b} \ge \vert \widehat{Q}^{e+b}_{02} \vert \ge
-\widehat{T}^{e+b}_{22}.
\label{inequality1} \ee
For sub-critical background $e^2 < (1 + b^2)$, the
fundamental string fluid behaves as a non-relativistic medium
in that the sound velocity of the fluid is given by $v_s$.
One readily finds that the two inequalities
Eqs.(\ref{inequality1}, \ref{inequality2}) are saturated precisely
in the critical limit $e^2 = (1 + b^2)$.

\section{Comparison with Effective Field Theory Approach}
Finally, we shall be considering the rolling tachyon with electric
and magnetic fields in the effective field theory approach.
Consider the D$p$-brane effective action of the form
\begin{eqnarray}
S_{\rm DBI} =-\cT_p \int_{\Sigma_{p+1}} \d^{p+1}x \,
\,V(T)\sqrt{-\det(\eta+F)}\,\cF(z), \label{lag}
\end{eqnarray}
where
\begin{eqnarray}
z=\l({1 \over  \eta+F}\r)^{(ab)}\partial_a T\partial_b T.
\nonumber
\end{eqnarray}
We will not make use of the explicit form of the function $\cF(z)$
and the tachyon potential $V(T)$ in the following discussion. If
we choose
\begin{eqnarray}
\cF(z)=\frac{z 4^z\Gamma(z)^2}{2\Gamma(2z)} \qquad {\rm and}
\qquad V(T)=e^{-\frac{1}{4}T^2}, \nonumber
\end{eqnarray}
the action Eq.(\ref{lag}) coincides with the
background-independent string field theory action for a non-BPS
D-brane in super-string theory given in \cite{KuMaMo,TeUe}. We can
also get the action proposed in \cite{Garousiaction} with an
alternative choice $\cF(z)=\sqrt{1+z}$. A similar treatment can be
found in \cite{GHY}.

We consider the situation considered in section \ref{elemagroll},
and assume that we can safely restrict our consideration with
homogeneous tachyon rolling with constant electric field
$e=F_{02}$ and magnetic field $b=F_{12}$. Then, the D-brane
world-volume Lagrangian density becomes
\begin{eqnarray}
{\cal L}= -\cT_p\,V(T)\sqrt{1-e^2+b^2}\,\cF(z), \nonumber
\end{eqnarray}
in which
\begin{eqnarray}
z=-\left(\frac{1+b^2}{1-e^2+b^2}\right)\,\dot T^2(t). \nonumber
\end{eqnarray}
The canonical conjugate momenta are
\begin{eqnarray}
\pi\equiv\frac{\delta\cL}{\delta e}
&=&\cT_p\,\frac{e}{\sqrt{1-e^2+b^2}}
V(T)\cD(z),
\label{pi}\\
P_T\equiv\frac{\delta\cL}{\delta \dot T}
&=&\cT_p\frac{2(1+b^2)}{\sqrt{1-e^2+b^2}}V(T)\cF'(z) \, \dot T,
\nonumber
\end{eqnarray}
where we have defined $\cD(z):=\cF(z)-2z\cF'(z)$. The Hamiltonian
density is then obtained as
\begin{eqnarray}
\cH=\Big( e\pi+\dot TP_T-\cL \Big)
=\cT_p\,\frac{1+b^2}{\sqrt{1-e^2+b^2}} V(T)\cD(z). \label{Hami}
\end{eqnarray}
This Hamiltonian density should be compared with the energy
density $T_{00}^{e+b}$ computed in the previous section from the
corresponding boundary state. We find a complete agreement between
Eq.(\ref{Teb}) and Eq.(\ref{Hami}) {\sl provided} we identify
\begin{eqnarray}
V(T)\cD(z) \quad\longleftrightarrow \quad \half\l(
1+\cos(2\pi\tilde\lambda) \r). \label{rel1}
\end{eqnarray}
For example, with $F_{ab} = \dot{T} = 0$, ${\cal D}(z) = 1$ and $
V(T) \rightarrow 0$ as $\widetilde{\lambda} \rightarrow 1/2$. This
explains Sen's identification of the closed string vacuum with
$\wt{\lambda} = 1/2$.

Coupling to the Kalb-Ramond field $B_{ab}$ can be introduced by
replacing $F_{ab}$ with the gauge-invariant combination
$(F_{ab}+B_{ab})$ in Eq.(\ref{lag}). As such, the source for
$B_{ab}$ field is obtained by varying the Lagrangian density with
respect to $F_{ab}$. Therefore, the electric displacement field
$\pi$ in Eq.(\ref{pi}) ought to correspond to
$Q^{02}_{e+b}=-Q_{02}^{e+b}$ in Eq.(\ref{Seb}). In fact, they
agree each other once the identification Eq.(\ref{rel1}) is made.
Similarly, the source current tensor that couples to the $B_{12}$
field is obtained as
\begin{eqnarray}
\frac{\del\cL}{\del b}=-\cT_p\l( \gamma \widetilde{\gamma} \left(
\frac{e^2 b}{1+b^2}\right) V(T) \cD(z) +{1 \over \gamma
\widetilde{\gamma}} \left(\frac{b}{1+b^2}\right) V(T)\cF(z)\r).
\nonumber
\end{eqnarray}
Again, this agrees with $Q_{12}^{e+b}$ in Eq.(\ref{Seb}) {\sl
provided}, in addition to the previous identification
Eq.(\ref{rel1}), we also make the identification
\begin{eqnarray}
V(T)\cF(z)\quad \longleftrightarrow \quad f(\gamma^{-1}y^0).
\label{rel2}
\end{eqnarray}

The energy-momentum tensor is computable from the Lagrangian
density Eq.(\ref{lag}) by replacing the flat metric $\eta_{ab}$
with a curved one $g_{ab}$ and keep the terms which are linear
with respect to the virtual variation $\delta
g_{ab}=g_{ab}-\eta_{ab}$. Some useful relations are
\begin{eqnarray}
\delta\sqrt{-\det(g+F)} &=&\sqrt{-\det(\eta+F)} \l({1 \over
\eta+F}\r)^{(ab)}\delta g_{ab}\nn &=&\gamma\wt\gamma \l(
{-(1+b^2)\delta g_{00} + be(\delta g_{01}+\delta g_{10})
+(1-e^2)\delta g_{11}+\delta g_{22} }\r) \nn
&+&\frac{1}{\gamma\wt\gamma} \sum_{i\ge3}\delta g_{ii}, \nonumber
\end{eqnarray}
and
\begin{eqnarray}
\delta z&=&-\l(\frac{1}{\eta+F}\,\delta
g\,\frac{1}{\eta+F}\r)^{(ab)} \del_a T\del_b T\nn &=& -\gamma^4
\widetilde{\gamma}^4 \l[ {(1+b^2)\delta g_{00}-be(1+b^2)(\delta
g_{01}+\delta g_{10}) +b^2 e^2 \delta g_{11}-e^2 \delta g_{22}}\r]
\dot T^2. \nonumber
\end{eqnarray}
As a result, we obtain
\begin{eqnarray}
\frac{\delta \cL}{\delta g_{00}}&=&\frac{\cT_p}{2} \gamma
\widetilde{\gamma} (1+b^2) \,V(T)\cD(z),\nn
\frac{\delta\cL}{\delta g_{11}}&=&\frac{\cT_p}{2}\l( +\gamma
\widetilde{\gamma} \frac{e^2b^2}{1+b^2}\,V(T)\cD(z) -{1 \over
\gamma \widetilde{\gamma}} \frac{1}{1+b^2}\,V(T)\cF(z)\r),\nn
\frac{\delta\cL}{\delta g_{22}}&=&\frac{\cT_p}{2} \l( -\gamma
\widetilde{\gamma} \frac{e^2}{1+b^2}\,V(T)\cD(z) -{1 \over \gamma
\widetilde{\gamma}} \frac{1}{1+b^2}\,V(T)\cF(z)\r),\nn
\frac{\delta \cL}{\delta g_{ii}}&=&-\frac{\cT_p}{2} {1 \over
\gamma \widetilde{\gamma}} \,V(T)\cF(z), \qquad \qquad (i = 3,
\cdots, p),\nn \frac{\delta \cL}{\delta g_{01}}&=&\frac{\delta
\cL}{\delta g_{10}} =-\frac{\cT_p}{2}\gamma \widetilde{\gamma}
\,eb\,V(T)\cD(z). \nonumber
\end{eqnarray}
We confirm that all these reproduce correctly the energy-momentum
tensor given in Eq.(\ref{Teb}), once we make use of the relations
Eq.(\ref{rel1}) and Eq.(\ref{rel2}).

The coupling to the sigma model dilaton field is obtainable by
multiplying $e^{-\varphi}$ factor to the D$p$-brane world-volume
Lagrangian density. As such, the dilaton tadpole is given by
\begin{eqnarray}
\rho_{\rm dilaton}=-\cL(T)= {\cal T}_p V(T)\sqrt{1-e^2+b^2}\cF(z).
\nonumber
\end{eqnarray}
This agrees with the result Eq.(\ref{tacdila}) obtained from the
boundary state formalism in the previous section {\sl provided}
the identification Eq.(\ref{rel2}) is again used.

How does the coupling to the closed string modes look like in the
Einstein frame? Take, for instance, the massless modes. The
dilaton, metric, and Kalb-Ramond fields in the Einstein frame,
$\varphi^E, g_{ab}^E, B_{ab}^E$, are related to those in the sigma
model frame as
\begin{eqnarray}
\varphi^E=\varphi, \qquad g^{E}_{ab}=e^{-\alpha\varphi}g_{ab},
\qquad B^{E}_{ab}=e^{-\alpha\varphi}B_{ab}, \nonumber
\end{eqnarray}
where $\alpha=1/6$ for bosonic string and $\alpha=1/2$ for
super-string theories, respectively. Expanding the D$p$-brane
world-volume Lagrangian density up to the linear order with
respect to the fluctuation of these fields around the flat
background, we obtain
\begin{eqnarray}
\delta \cL &\sim& \rho_{\rm dilaton}\delta\varphi +\half
T^{ab}\delta g_{ab}
 +\half Q^{ab}\delta B_{ab} \nonumber \\
&\sim& \left( \rho_{\rm dilaton}
+\frac{\alpha}{2}\eta_{ab}T^{ab}\right) \delta\varphi^E +\half
T^{ab}\delta g^E_{ab}
 +\half Q^{ab}\delta B^E_{ab}. \nonumber
\end{eqnarray}
It shows that the energy-momentum tensor and the string current
tensor remains unchanged, while the dilaton coupling is modified
in the Einstein frame to:
\begin{eqnarray}
\rho_{\rm dilaton}^E= \rho_{\rm
dilaton}+\frac{\alpha}{2}\eta_{ab}T^{ab}. \nonumber
\end{eqnarray}
As the trace of the energy-momentum tensor asymptotes to a
non-vanishing constant value at late time (see Eq.(\ref{Teb})), we
see that the dilaton coupling remains finite in the Einstein
frame, though it vanishes in the sigma model frame.

\section*{Acknowledgement}
SJR thanks Ik-Jae Shin, Fumihiko Sugino and  Sangheon Yoon for
useful discussions, and Paolo Di Vecchia and staffs at NORDITA and
Niels Bohr Institute for warm hospitality during this work. SS
thanks Koji Hashimoto, Takuya Okuda, Ashoke Sen, Seiji Terashima,
Paolo Di Vecchia and Takashi Yokono for useful discussions, and
String Theory group at Seoul National University for warm
hospitality during this work. The work of SJR is supported in part
by the Fellowship of the Institute for Advanced Study. The work of
SS is supported in part by Danish Natural Science Research
Council.

\end{document}